\documentclass{article}
\usepackage{spconf,amsmath,graphicx}
\usepackage{amsmath,graphicx}
\usepackage{color}
\usepackage{amssymb}
\usepackage{booktabs}
\usepackage{hyperref}
\usepackage{multirow}
\usepackage{comment}

\title{HiFi-WaveGAN: Generative Adversarial Network with Auxiliary Spectrogram-Phase Loss for High-Fidelity Singing Voice Generation}
\name{$^*${Chunhui Wang}$^1$, $^*${Chang Zeng}$^{2,3}$, Jun Chen$^4$, Xing He$^1$}
\address{$^1$Beijing Bombax XiaoIce Technology Co., Ltd, China \\ $^2$National Institute of Informatics, Japan $^3$SOKENDAI, Japan \\ $^4$Shenzhen International Graduate School, Tsinghua University, Shenzhen, China
}
\begin{document}
\ninept
\maketitle
\renewcommand{\thefootnote}{\fnsymbol{footnote}}
\footnotetext[1]{These authors contributed equally to this work.}
%
\begin{abstract}
Entertainment-oriented singing voice synthesis (SVS) requires a vocoder to generate high-fidelity (e.g. $48$kHz) audio. However, most text-to-speech (TTS) vocoders cannot reconstruct the waveform well in this scenario. In this paper, we propose HiFi-WaveGAN to synthesize the $48$kHz high-quality singing voices in real-time. Specifically, it consists of an Extended WaveNet served as a generator, a multi-period discriminator proposed in HiFiGAN, and a multi-resolution spectrogram discriminator borrowed from UnivNet. To better reconstruct the high-frequency part from the full-band mel-spectrogram, we incorporate a pulse extractor to generate the constraint for the synthesized waveform. Additionally, an auxiliary spectrogram-phase loss is utilized to approximate the real distribution further. The experimental results show that our proposed HiFi-WaveGAN obtains $4.23$ in the mean opinion score (MOS) metric for the $48$kHz SVS task, significantly outperforming other neural vocoders.
\end{abstract}
\section{Introduction}
\label{sec:intro}
Speech synthesis has been a prominent area of research in the speech community for an extended period. With the advent of deep learning, neural networks have replaced various components in the speech synthesis pipeline, including front-end text analysis \cite{front1, front2}, acoustic model \cite{tacotron2, fastspeech, gantts, wgantts, naturaltts}, and vocoder \cite{pwg, hifigan, melgan}. Singing voice synthesis (SVS), given its similarity to speech synthesis in workflow, can benefit significantly from the advancements in Text-to-Speech (TTS) \cite{earlysvs, s2sfft, xiaoicesing, wgan}.

For instance, \cite{xiaoicesing} proposed an acoustic model for SVS based on Fastspeech \cite{fastspeech, fastspeech2}, originally designed for speech synthesis. However, when it comes to the vocoder, 
most studies \cite{xiaoicesing, bytesing} have simply borrowed existing models such as WORLD \cite{world} and WaveRNN \cite{wavernn} from the speech synthesis domain without customizing them to account for the unique characteristics of singing voices. Consequently, these vocoders struggle to accurately reconstruct the high-frequency components of the mel-spectrogram, particularly when dealing with entertainment-oriented scenarios that require generating high-fidelity (e.g., 48kHz) singing voices.

The SingGAN vocoder \cite{singgan} designed for SVS highlighted several reasons why vanilla TTS vocoders are not well-suited for generating natural singing voices, including:
\begin{itemize}
\setlength{\itemindent}{.2in}
\item Long continuous pronunciation,
\item Strong expressiveness,
\item Higher sampling rate than speech.
\end{itemize}
SingGAN addressed these characteristics by designing a generative adversarial network capable of generating singing voices at a $24$kHz sampling rate. However, this sampling rate limitation may impact the quality of the singing voices, particularly in the high-frequency parts, as it cannot fully capture information beyond $12$kHz, in accordance with Nyquist's sampling law. Therefore, there is room to enhance the quality further by generating singing voices at a higher sampling rate.

To address this, we propose a novel HiFi-WaveGAN in this paper, capable of generating high-fidelity singing voices at a human-level quality with a 48kHz sampling rate. Our approach consists of an Extended WaveNet \cite{wavenet} (ExWaveNet) serving as the generator, a multi-period discriminator (MPD) proposed in HiFiGAN \cite{hifigan}, and a multi-resolution spectrogram discriminator (MRSD) borrowed from UnivNet \cite{univnet}.

For the ExWaveNet generator, we expand the kernel size of the convolutional layers to accommodate a larger receptive field, enabling it to handle the long continuous pronunciation in SVS. Recognizing that pitch carries greater expressiveness in SVS compared to TTS, we concatenate it with the mel-spectrogram and input it into an upsample network.

To address the challenge of accurately reconstructing the high-frequency regions from the full-band mel-spectrogram, we introduce an additional Pulse Extractor (PE) to generate a pulse sequence, which is then incorporated as a conditional constraint input to the ExWaveNet. This operation plays a crucial role in rectifying unexpected distortions in the waveform. Furthermore, we leverage the auxiliary spectrogram-phase loss \cite{nsf}, which is combined with the adversarial training loss and feature match loss, to supervise the training process of our HiFi-WaveGAN. The inclusion of the phase-related term in the auxiliary loss enhances our model's ability to approximate the real distribution more effectively when compared to models that lack this component.

In the experiment, Xiaoicesing2 \cite{xiaoice2} is combined with the proposed HiFi-WaveGAN and other neural vocoders, which are used as the baseline models. The experimental results show that the quality of the synthesized singing voice of our proposed HiFi-WaveGAN not only outperforms all baselines but also comes very close to the human level under the MOS metric.

The rest of this paper is organized as follows. HiFi-WaveGAN is illustrated in Section \ref{sec:proposed} in detail, including the ExWaveNet generator, discriminators, and training loss functions. The experimental settings and results\footnote{\href{https://wavelandspeech.github.io/hifi-wavegan/}{\text{Demo page: https://wavelandspeech.github.io/hifi-wavegan/}}} are reported in Section \ref{sec:exp}. Finally, we conclude our paper in Section \ref{sec:con}. 

\begin{figure}[t]
  \centering
  \includegraphics[width=8cm]{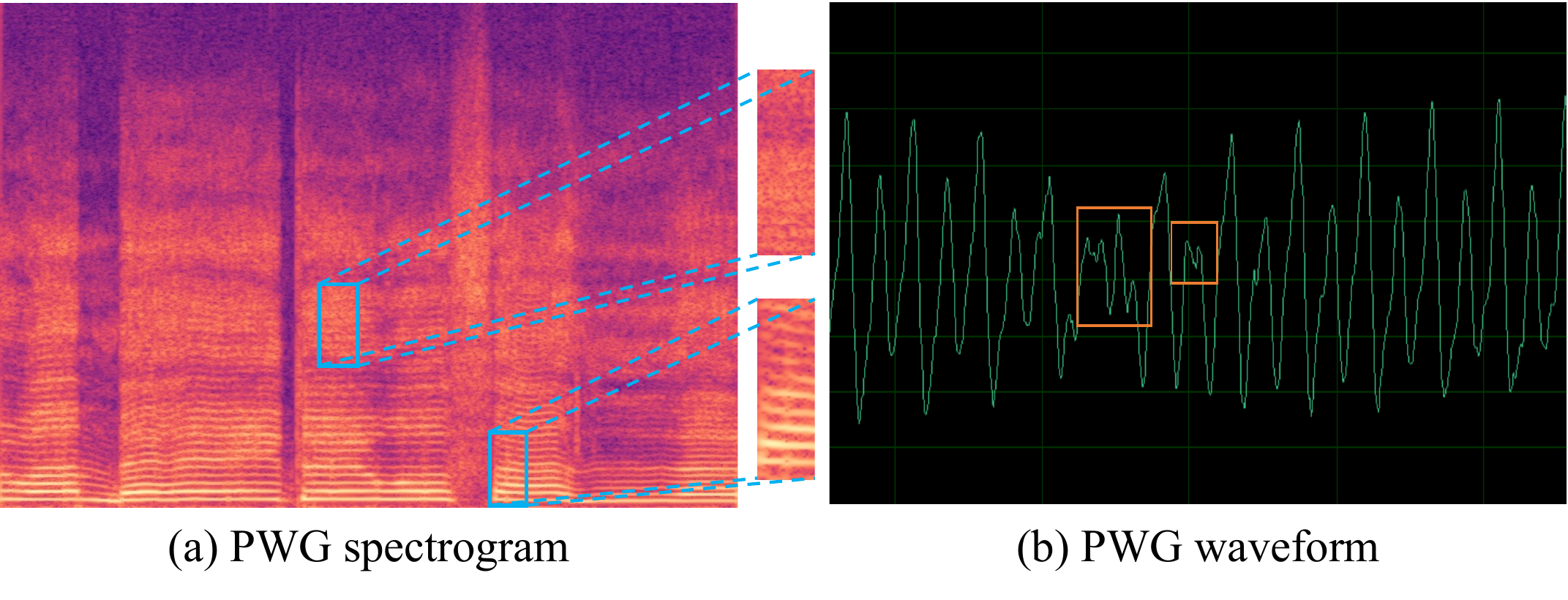}
  \caption{Spectrogram and waveform generated by PWG.}
  \label{fig:pwg}
\end{figure}

\section{HiFi-WaveGAN}
\label{sec:proposed}
Before introducing our proposed HiFi-WaveGAN, we conducted an analysis of the limitations of TTS neural vocoders when synthesizing $48$kHz singing voices, using the PWG vocoder as an example. In Figure \ref{fig:pwg} (a), we observed glitches in the low-frequency part that disrupt the continuity of the spectrogram. These glitches arise because the receptive field of PWG is insufficient to cover the long continuous pronunciation required for singing voices. Furthermore, the high-frequency harmonics in Figure \ref{fig:pwg} (a) appear blurry, indicating that the PWG neural vocoder struggles to accurately reconstruct the high-frequency components necessary for SVS tasks. We further investigated the source of these problems in the PWG spectrogram, as well as in other neural TTS vocoders. Figure \ref{fig:pwg} (b) reveals periodic distortions in the waveform generated by the PWG vocoder, leading to trembling and low-quality singing voices.

To address the aforementioned issues, we present a detailed illustration of our HiFi-WaveGAN in the subsequent part of this section. As depicted in Figure \ref{fig:arch-hwg}, our HiFi-WaveGAN consists of an ExWaveNet generator, responsible for generating high-quality singing voices. Additionally, we employ two independent discriminators: MPD (Multi-Period Discriminator) and MRSD (Multi-Resolution Spectrogram Discriminator). These discriminators are designed to distinguish real/fake waveforms from periodic patterns and consecutive long dependencies, respectively.


\subsection{Extended WaveNet}
Similar to the generator in PWG \cite{pwg}, we also adopt a WaveNet-based model as the generator. However, recognizing that singing voices exhibit longer continuous pronunciation compared to speech \cite{singgan}, we enhance the architecture of WaveNet by utilizing an $18$-layer one-dimensional CNN with larger kernel sizes, resulting in an improved model called \textbf{Extended WaveNet} (ExWaveNet). Specifically, we evenly divide the $18$ layers of the generator into three stacks, and the kernel sizes in each stack are set to $\{3,3,9,9,17,17\}$, determined through neural architecture search \cite{nas}. This modification empowers the network to effectively capture longer continuous pronunciation in singing voices with high sampling rates, courtesy of the increased receptive field.

In addition to modeling long continuous pronunciation, restoring expressiveness in singing voices is also crucial. To achieve this, we concatenate the pitch with the mel-spectrogram as the input to the upsampling network, following the same approach as in PWG. The upsampled representation is then concatenated with a pulse sequence $\boldsymbol{T}$, which will be explained in detail in the next paragraph, serving as the conditional input for generating singing voices with strong expressiveness. Moreover, we observed that upsampling the random input noise using an identical network also improves the expressiveness of the synthesized audio.

\begin{figure*}[t]
  \centering
  \includegraphics[width=12cm]{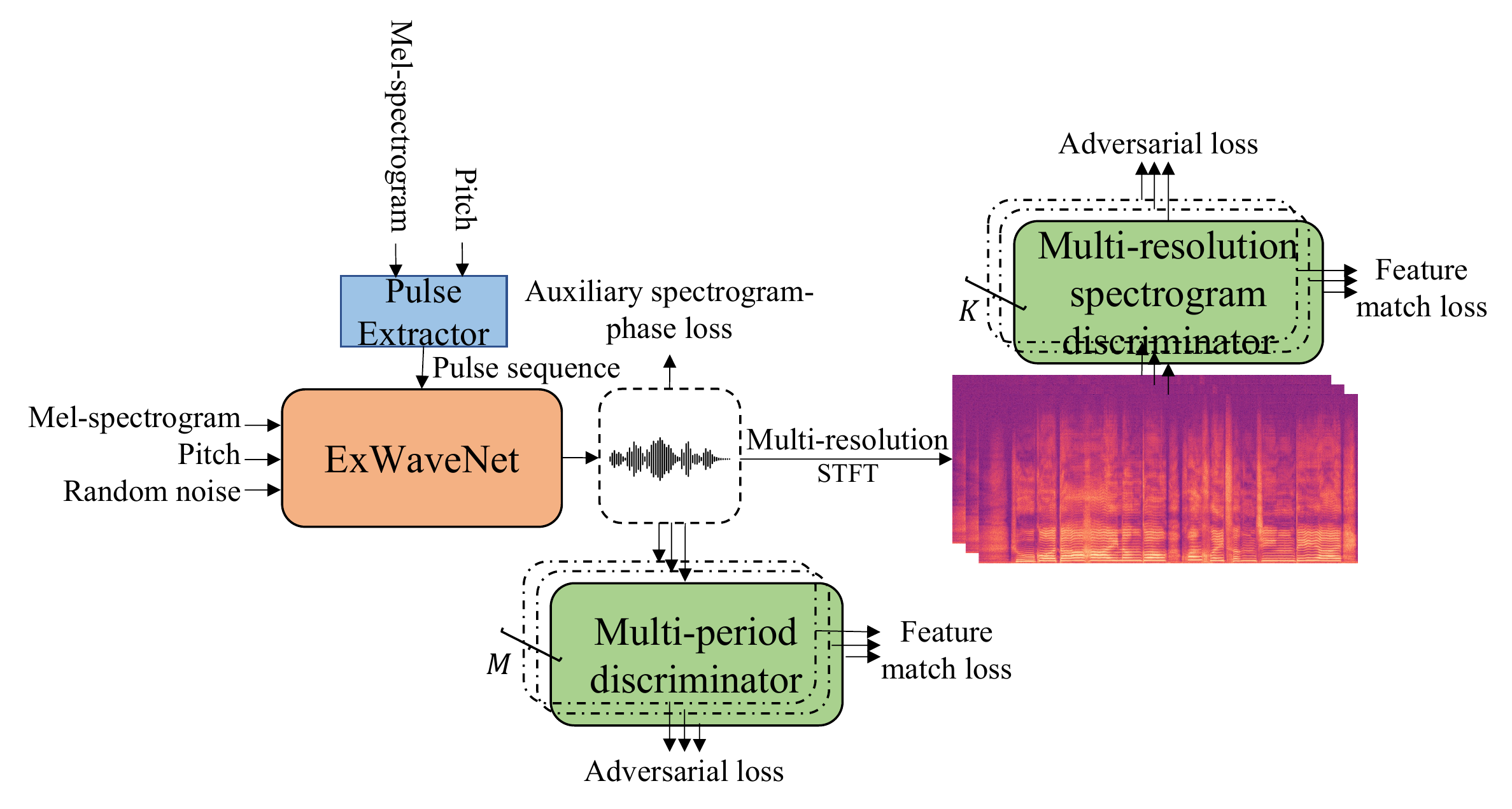}
  \caption{Architecture of proposed HiFi-WaveGAN. It consists of a generator ExWaveNet and two independent discriminators MPD and MRSD.}
  \label{fig:arch-hwg}
\end{figure*}

As mentioned in the previous paragraph, there are some periodic distortions in the waveform generated by TTS neural vocoders, leading to low-quality synthesized singing voices. To address this, we propose an additional \textbf{Pulse Extractor} (PE) to generate a pulse sequence as a constraint condition during waveform synthesis. As shown in Fig. \ref{fig:arch-hwg}, the extractor takes the mel-spectrogram and pitch as input. The pulse is extracted at each extreme point of the waveform envelope, determined by the V/UV decision and mel-spectrogram, and can be formulated as:

\begin{align}
\boldsymbol{T}[\ i \ ] =
\begin{cases}
||\boldsymbol{M}[\ i \ ]||_{F}, & UV = 1, i = \frac{s}{f_0} \\
0, & UV = 1, i \neq \frac{s}{f_0} \\
\text{noise}, & UV = 0
\end{cases}
\end{align}

where $\boldsymbol{M}$ represents the mel-spectrogram, $i$ is the time index, $||\ \cdot\ ||_F$ denotes the Frobenius norm, $\boldsymbol{T}[\ i\ ]$ indicates the pulse value at index $i$, and $s$ and $f_0$ denote the sampling rate and pitch, respectively. The noise in the formula is generated from a Gaussian distribution.

\subsection{Discriminators}
Identifying both consecutive long-term dependencies and periodic patterns plays a crucial role in modeling realistic audio \cite{hifigan}. In the proposed HiFi-WaveGAN, we employ two independent discriminators to evaluate singing voices from these two aspects.

The first discriminator is MRSD, adapted from UnivNet \cite{univnet}, which identifies consecutive long-term dependencies in singing voices from the spectrogram. We transform both real and fake singing voices into spectrograms using different combinations of FFT size, window length, and shift size. Then, two-dimensional convolutional layers are applied to the spectrograms. As depicted in Fig. \ref{fig:arch-hwg}, the model employs $K$ sub-discriminators, each utilizing a specific combination of spectrogram inputs. In our implementation, $K$ is set to four.

The second discriminator is MPD, identical to the one used in HiFiGAN \cite{hifigan}. It transforms the one-dimensional waveform with length $T$ into $2$-d data with height $T/p$ and width $p$ by setting the periods $p$ to $M$ different values, resulting in $M$ independent sub-discriminators within MPD. In this paper, we set $M$ to five and $p$ to $[2,3,5,7,11]$. As described in \cite{hifigan}, this design allows the discriminator to capture distinct implicit structures by examining different parts of the input audio.



\subsection{Loss function}
Similar to other models \cite{hifigan, singgan}, we adopt a weighted combination of multiple loss terms as the final loss function formulated by Eq. (\ref{eq:final_dloss}) and Eq. (\ref{eq:final_gloss}) to supervise the training process of our HiFi-WaveGAN. 
\begin{align}
    \mathcal{L}_D & = \mathcal{L}_{adv}(D;G), \label{eq:final_dloss} \\
    \mathcal{L}_G & = \lambda_1 * \mathcal{L}_{adv}(G;D) + \lambda_2 * \mathcal{L}_{aux} + \lambda_3 * \mathcal{L}_{fm}, \label{eq:final_gloss}
\end{align}
where $\mathcal{L}_{adv}$, $\mathcal{L}_{aux}$, and $\mathcal{L}_{fm}$ denote adversarial loss, auxiliary spectrogram-phase loss, and feature match loss, respectively. In this paper, $\lambda_1$, $\lambda_2$, and $\lambda_3$ are set to $1$, $120$, and $10$, respectively.

\subsubsection{Adversarial loss}
For the adversarial loss, we adopt the format in LS-GAN \cite{lsgan} to avoid the gradient vanishing. The formula is shown as
\begin{footnotesize}
\begin{align}
    \label{eq:adv_loss}
    \mathcal{L}_{adv}(G;D) & = \mathbb{E}_{\boldsymbol{z} \sim \mathcal{N}(0,1)}[(1 - D(G(\boldsymbol{z})))^2], \\
    \mathcal{L}_{adv}(D;G) & = \mathbb{E}_{\boldsymbol{y} \sim p_{data}}[(1 - D(\boldsymbol{y}))^2] + \mathbb{E}_{\boldsymbol{z} \sim \mathcal{N}(0,1)}[D(G(\boldsymbol{z}))^2],
\end{align}
\end{footnotesize}
where $G$ and $D$ denote the generator and discriminators, respectively, $\boldsymbol{z}$ is the random noise and $\boldsymbol{y}$ represents the real singing voice.

\subsubsection{Auxiliary spectrogram-phase loss}
The auxiliary spectrogram-phase loss is utilized to constrain the behavior of the generator by measuring the metrics between real and fake features including STFT spectrogram, STFT phase, and mel-spectrogram. First, we define two functions for calculating the convergence loss term and magnitude loss term \cite{scmag} as
\begin{align}
    \label{eq:sc_mag}
    \mathcal{L}_{c}(\boldsymbol{x},\boldsymbol{y}) & = \frac{||\boldsymbol{x} - \boldsymbol{y}||_F}{||\boldsymbol{x}||_F}, \\
    \mathcal{L}_{mag} & = \frac{1}{N}||\log \boldsymbol{x} - \log \boldsymbol{y}||_1,
\end{align}
where $||\ \cdot\ ||_1$ represents $L1$ norm, $\boldsymbol{x}$ and $\boldsymbol{y}$ respectively denote the fake and real features, and $N$ denotes the number of elements in the magnitude. Next, the above formulas are applied to STFT and mel-spectrogram as
\begin{footnotesize}
\begin{align}
    \label{eq:sc_mag_mel_stft}
    \mathcal{L}_{sp} = \frac{1}{H_1} \sum_{h=1}^{H_1}(& \mathcal{L}_{s\_c}^{(h)}(\boldsymbol{x}_{s}, \boldsymbol{y}_{s}) + \mathcal{L}_{s\_mag}^{(h)}(\boldsymbol{x}_{s}, \boldsymbol{y}_{s}) \nonumber \\ 
    & + \mathcal{L}_{p\_c}^{(h)} (\boldsymbol{x}_{p},\boldsymbol{y}_{p})), \\
    \mathcal{L}_{mel} = \frac{1}{H_2} \sum_{h=1}^{H_2}(& \mathcal{L}_{mel\_c}^{(h)}(\boldsymbol{x}_{mel}, \boldsymbol{y}_{mel}) + \mathcal{L}_{mel\_mag}^{(h)}(\boldsymbol{x}_{mel}, \boldsymbol{y}_{mel})),
\end{align}
\end{footnotesize}
where $H_1$ and $H_2$ are the numbers of different combinations of transformation parameters for STFT and mel-spectrogram, and they are set as $3$ and $2$ in this paper, respectively. Compared with the auxiliary spectrogram loss $\mathcal{L}_{s\_c}$ described in SingGAN \cite{singgan}, we use an additional phase convergence term $\mathcal{L}_{p\_c}$ for faster convergence. Finally, the auxiliary loss is defined as a summary of the mel-spectrogram term and STFT term as
\begin{align}
    \mathcal{L}_{aux} & = \mathcal{L}_{sp} + \mathcal{L}_{mel}
    \label{eq:aux_loss}
\end{align}

\subsubsection{Feature match loss}
Feature match loss was proposed in MelGAN \cite{melgan} to measure the $L1$ norm between the feature maps of real and fake audios, which are extracted from the hidden layers of the discriminator. In HiFi-WaveGAN, we apply the constraint to both MRSD and MPD as Fig. \ref{fig:arch-hwg} shows. Therefore, it can be defined as
\begin{align}
    \label{eq:fm_loss}
    \mathcal{L}_{fm} = \mathbb{E}_{\boldsymbol{z}, \boldsymbol{y}}[ & \sum_{i=1}^{L_{MRSD}}\frac{1}{N_i}||D^i(\boldsymbol{y})-D^i(G(\boldsymbol{z}))||_1 \nonumber \\
    & + \sum_{j}^{L_{MPD}}\frac{1}{N_j}||D^j(\boldsymbol{y})-D^j(G(\boldsymbol{z}))||_1],
\end{align}
where $L_{MRSD}$ and $L_{MPD}$ are the number of layers of MRSD and MPD, respectively, $D^i(\cdot)$ and $N_i$ represent the feature map and the number of feature map of the $i$-th layer in MRSD, and $D^j(\cdot)$ and $N_j$ the feature map and the number of feature map of the $j$-th layer of MPD. 

\section{Experiment}
\label{sec:exp}

\subsection{Dataset}
All experiments related to $48$kHz SVS were conducted on our internal singing dataset, which comprises $6917$ pieces sung by a female singer, with durations ranging from $4$ seconds to $10$ seconds. For the experiments, we randomly selected $300$ pieces as validation data and another $300$ pieces as testing data, while the remaining data were utilized for training. When preparing the acoustic features, we used a window length and shift length of $20$ms and $5$ms, respectively, for the Short-Time Fourier Transform (STFT). Furthermore, we applied $120$ mel filters to transform the spectrogram into mel-scale and subsequently normalized it. Additionally, pitch and V/UV (voiced/unvoiced) decisions were also extracted from the data.

\subsection{Experimental setup}
To evaluate the performance of our proposed HiFi-WaveGAN and make a comparative analysis with other vocoders in the context of $48$kHz SVS, we integrated them with an acoustic model named Xiaoicesing2 \cite{xiaoice2}. The acoustic model, based on Fastspeech, comprises an encoder, a duration predictor, and a decoder, collectively forming a comprehensive SVS system. The input to this SVS system consists of the musical score, encompassing lyrics, note pitch, and note duration. These input data are transformed into corresponding phoneme embeddings, pitch embeddings, and duration embeddings. The encoder then combines these embeddings and transforms the sequence into a dense vector sequence in a hidden semantic space. As seen in Fastspeech \cite{fastspeech}, a length regulator is employed to expand the length of the dense vector sequence based on the predicted phoneme duration from the duration predictor. Finally, the decoder generates mel-spectrograms, V/UV decisions, and logarithmic F0 (logF0) for all the vocoders under consideration.

The training of the acoustic model is carried out for $40$k iterations, employing a batch size of $32$, and employing the Adam \cite{adam} optimizer with hyperparameters $\beta_1=0.9$, $\beta_2=0.98$, and $\epsilon=10^{-9}$. This configuration facilitates an effective and comprehensive evaluation and comparison of the performance of HiFi-WaveGAN and other vocoders within the $48$kHz SVS task.

In our experiment, we selected three neural vocoders, namely Parallel WaveGAN \cite{pwg}, HiFiGAN \cite{hifigan}, and RefineGAN \cite{refinegan}, as the baseline models for high-fidelity singing voice generation. These vocoders were combined with the acoustic model to form complete SVS systems. During training, we utilized the ground truth mel-spectrogram, V/UV decision, and logF0 as inputs to train these vocoders. When synthesizing the high-fidelity waveform, the acoustic model generated the mel-spectrogram, V/UV decision, and logF0, which were then used as input features for the vocoders.

It is worth noting that PWG and HiFiGAN vocoders were originally designed for speech synthesis with a sampling rate of $16$kHz or $22$kHz. To enable a fair comparison, we also improve the architecture of the generator of them to generate singing voices at a higher sampling rate of $48$kHz. As a result, we referred to these adapted versions as PWG-$48$kHz and HiFiGAN-$48$kHz in our experiments. This adaptation allows these vocoders to generate high-fidelity singing voices with the desired sampling rate for evaluation and comparison.


\subsection{Training methodology}
To train the HiFi-WaveGAN, we randomly choose $4$s as a training sample and train the model for $200$k iterations with $8$ batch size by using an AdamW \cite{adamw} optimizer with $0.0002$ learning rate and $0.01$ weight decay. Other hyper-parameters of AdamW are set as $\beta_1=0.8$ and $\beta_2=0.99$. An exponential learning rate scheduler with $0.999$ learning rate decay is used to better optimize the model. The training process costs about $70$ hours on $4$ NVIDIA V100 GPUs.

\subsection{Experimental result}
\begin{table}[t]
\footnotesize
  \caption{Subjective and objective test result of different vocoders for 48kHz singing voice synthesis.}
  \vspace{4mm}
  \setlength\tabcolsep{1.5pt}
  \label{tab:result_mos_svs}
  \centering
  \begin{tabular}{lrcccc}
    \toprule
    \textbf{Vocoder} & \textbf{\#Params}($\downarrow$) & \textbf{RTF}($\downarrow$) & \textbf{MOS}($\uparrow$) & \textbf{STOI}($\uparrow$) & \textbf{PESQ}($\uparrow$) \\
    \midrule
    Ground truth           & -                    & -                    & $4.27$              & -                     & -           \\
    PWG \cite{pwg}         & $\boldsymbol{1.54}$M & $0.027$              & $2.55$              & $0.8406$              & $3.40$      \\
    PWG-$48$kHz & $9.83$M & $0.028$ & $2.78$ & $0.8894$ & $3.57$ \\
    HiFiGAN \cite{hifigan} & $14.47$M             & $\boldsymbol{0.014}$ & $3.77$              & $0.9403$              & $3.87$      \\
    HiFiGAN-$48$kHz & $19.90$M & $0.025$ & $3.89$ & $0.9422$ & $3.95$ \\
    RefineGAN \cite{refinegan} & $7.68$M & $0.034$ & $4.15$ & $0.9432$ & $4.01$ \\
    HiFi-WaveGAN                  & $10.08$M             & $0.026$              & $\boldsymbol{4.23}$ & $\boldsymbol{0.9524}$ & $\boldsymbol{4.03}$          \\
    \bottomrule
  \end{tabular}
\end{table}

As our focus is on $48$kHz singing voice generation, the number of parameters for PWG and HiFiGAN in Table \ref{tab:result_mos_svs} appears slightly larger compared to the numbers reported in \cite{pwg} and \cite{hifigan}. This is due to our adaptation of these models to handle the higher sampling rate. Specifically, our HiFi-WaveGAN has more parameters than PWG, as we utilize larger kernel sizes to model long continuous pronunciation. Despite this, the inference speed of HiFi-WaveGAN remains comparable to PWG, as demonstrated in the table. HiFiGAN, on the other hand, boasts the fastest inference speed in this comparison, but it comes with the highest number of parameters. Meanwhile, RefineGAN \cite{refinegan} has fewer parameters than our HiFi-WaveGAN but has the slowest inference speed among the models in the table.

For evaluating the quality of the synthesized audio, we conducted both subjective and objective assessments comparing our HiFi-WaveGAN with other vocoders, as well as the ground truth. The subjective evaluation involved preparing $20$ segments of singing voices for each vocoder and ground truth, with durations ranging from $6$s to $10$s. Twenty listeners participated in our listening test, and the results in Table \ref{tab:result_mos_svs} reveal that our HiFi-WaveGAN obtained the highest scores among all vocoders. Notably, despite both PWG and HiFi-WaveGAN utilizing WaveNet as the generator, our HiFi-WaveGAN significantly outperformed PWG by $1.68$ and PWG-$48$kHz by $1.45$ in terms of MOS. This improvement can be attributed to the adoption of larger kernel sizes and the addition of the extra Pulse Extractor in our HiFi-WaveGAN.

Moreover, in comparison to HiFiGAN, HiFi-WaveGAN achieved a MOS score that was $0.46$ higher, primarily due to better reconstruction of high-frequency parts. Although HiFiGAN-$48$kHz achieved a slightly higher MOS score than HiFiGAN, it still did not surpass the performance of HiFi-WaveGAN. Interestingly, the subjective result of our HiFi-WaveGAN even outperformed RefineGAN, which was specifically designed for generating high-quality singing voices (at $44.1$kHz in \cite{refinegan}). It is worth noting that the MOS score of HiFi-WaveGAN was competitive with the ground truth, demonstrating its high quality close to the human level.

For objective evaluation, we calculated STOI and PESQ scores for all vocoders, and the results, shown in Table \ref{tab:result_mos_svs}, indicate that our HiFi-WaveGAN achieved the best performance in these metrics as well. This further demonstrates the superior quality of HiFi-WaveGAN compared to the other vocoders in the context of $48$kHz SVS.

\subsection{Spectrogram analysis}

\begin{figure}[t]
  \centering
  \includegraphics[width=8cm]{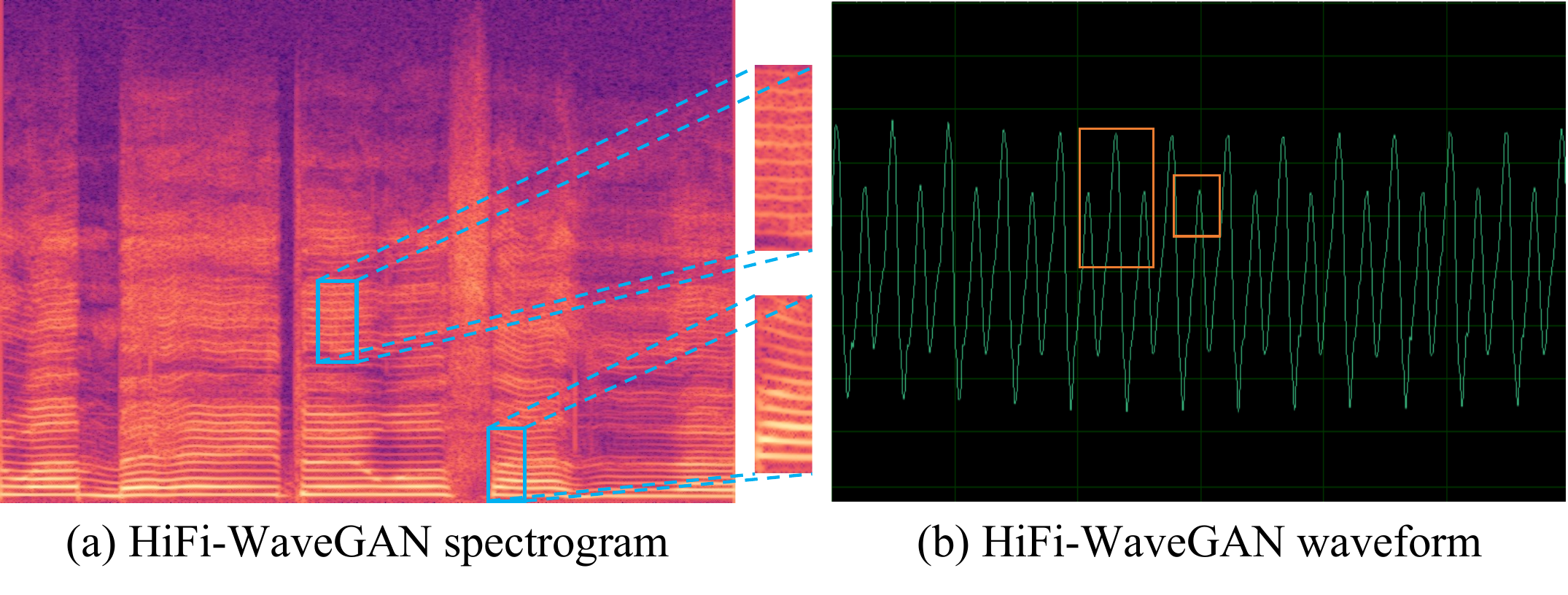}
  \caption{Spectrogram and waveform generated by HiFi-WaveGAN.}
  \label{fig:hwg}
\end{figure}

In this section, we conduct a detailed comparison of the waveforms and spectrograms generated by our HiFi-WaveGAN and Parallel WaveGAN to gain deeper insights. Starting with the waveforms shown in Fig. \ref{fig:pwg} (b) and Fig. \ref{fig:hwg} (b), it is evident that the distortions present in the orange solid box of the waveform generated by Parallel WaveGAN have been effectively rectified in our HiFi-WaveGAN. This outcome confirms that the pulse sequence generated by the Pulse Extractor plays a crucial role in addressing the defects found in the waveform generated by PWG. Furthermore, upon observing the low-frequency part of the spectrogram in Fig. \ref{fig:pwg} (a) and Fig. \ref{fig:hwg} (a), it becomes apparent that the glitches, which correspond to the distortions in the waveform, have also been successfully eliminated from the spectrogram in HiFi-WaveGAN. This achievement is primarily attributed to the pulse sequence and the larger receptive field of our generator.

Moreover, a comparison of the high-frequency part of the spectrogram in Fig. \ref{fig:pwg} (a) and Fig. \ref{fig:hwg} (a) reveals that the harmonics in Fig. \ref{fig:hwg} (a) are notably clearer than those in Fig. \ref{fig:pwg} (a). This improvement indicates that our HiFi-WaveGAN excels at reconstructing the rich details of the high-frequency components. This success can be attributed to our approach of incorporating informative pitch information into the mel-spectrogram and combining it with the pulse sequence to serve as the conditional constraint input to the generator.

Overall, this thorough analysis highlights the superior performance of our HiFi-WaveGAN, which effectively addresses waveform distortions and achieves high-quality reconstruction of both low and high-frequency components in the generated singing voices. The incorporation of the pulse sequence and the use of pitch information prove to be instrumental in enhancing the fidelity and expressiveness of the synthesized audio.


\begin{table}[t]
  \caption{Subjective and objective test result of HiFi-WaveGAN ablation study.}
  \vspace{4mm}
  \setlength\tabcolsep{5pt}
  \label{tab:result_ablation}
  \centering
  \begin{tabular}{lccc}
    \toprule
    \textbf{Vocoder} & \textbf{MOS}($\uparrow$) & \textbf{STOI}($\uparrow$) & \textbf{PESQ}($\uparrow$) \\
    \midrule
    Ground truth           & $4.27$              & -                     & -           \\
    HiFi-WaveGAN           & $\boldsymbol{4.23}$ & $\boldsymbol{0.9524}$ & $\boldsymbol{4.03}$          \\
    w/o ExWaveNet          & $3.89$              & $0.9404$              & $3.88$ \\
    w/o Pulse Extractor                & $3.10$              & $0.9069$              & $3.65$ \\
    w/o Phase loss         & $4.18$              &  $0.9408$             & $3.91$ \\
    \bottomrule
  \end{tabular}
\end{table}

\subsection{Ablation study}
Since we proposed three key components to synthesize high-quality $48$kHz singing voices, we conducted an ablation study to determine the individual contribution of each component, as shown in Table \ref{tab:result_ablation}.

First, we compared the results of HiFi-WaveGAN with the third row, where ExWaveNet was replaced with the original WaveNet. In this configuration, both subjective and objective metrics experienced a degradation, showcasing that the original WaveNet was inadequate in modeling the long pronunciation present in singing voices at this sampling rate.

Next, we examined the fourth row, where we removed the Pulse Extractor from our HiFi-WaveGAN. The significant decrease in the MOS score compared to the HiFi-WaveGAN result demonstrates that the Pulse Extractor is a pivotal component for synthesizing high-quality singing voices in our model.

Finally, in the fifth row, we omitted the phase loss term. Although MOS and other objective metrics only exhibited a slight decline compared to the HiFi-WaveGAN result, it can prove that the phase loss term leads to our model generating singing voices that were even closer to the human level in terms of quality.

Overall, the ablation study confirms the importance of each proposed component in our HiFi-WaveGAN. The ExWaveNet's ability to model long pronunciation, the Pulse Extractor's role in correcting waveform distortions, and the impact of the phase loss term on achieving high-quality singing voices all contribute significantly to the success of our model.

\section{Conclusion}
\label{sec:con}
In this paper, we present HiFi-WaveGAN, a novel approach designed to generate high-quality $48$kHz singing voices. Our method involves customizing the architecture of WaveNet to enhance its ability to model long continuous pronunciation, a crucial aspect of singing. Additionally, we introduce a novel Pulse Extractor, which is integrated with the customized WaveNet to effectively address waveform distortions, reduce glitches, and improve the clarity of high-frequency harmonics.

Furthermore, we incorporate pitch information into the mel-spectrogram as a condition, coupled with the upsampled intermediate representation from random input noise. This integration enhances the expressiveness of high-frequency parts in the synthesized singing voices. The results of our MOS test demonstrate that the quality of singing voices generated by HiFi-WaveGAN is remarkably close to human-level perception.

In the future, our work will concentrate on enhancing the acoustic model for $48$kHz SVS and combining it with HiFi-WaveGAN. This collaboration will further contribute to the advancement of high-fidelity singing voice synthesis and enable even more realistic and expressive singing voices to be created.

\bibliographystyle{IEEEbib}
\bibliography{strings,refs}

\end{document}